\documentclass{article}
\usepackage{smc2026}

\usepackage[english]{babel}
\usepackage[caption=false, font=footnotesize]{subfig}
\usepackage{paralist}
\usepackage{array}
\usepackage{multirow}
\usepackage{bm}
\usepackage{booktabs}
\usepackage[figure,table]{hypcap}

\newcommand{\papertitle}{Score-Informed Transformer for Refining MIDI Velocity in Automatic Piano Transcription}
\title{\papertitle}

\author[1]{\mbox{\firstname{Zhanhong}\lastname{He}}\email{zhanh.he.uwa@gmail.com}}
\author[1]{\mbox{\firstname{Roberto}\lastname{Togneri}}}
\author[1]{\mbox{\firstname{Defeng (David)}\lastname{Huang}}}

\affil[1]{\department{Signal Processing and Recognition Lab}\institution{University of Western Australia}\city{Perth}\state{WA}\postcode{6000}\country{Australia}\affiliationtype{University}}

\completesetup
\hypersetup{
  pdftitle  = {Score-Informed Transformer for Refining MIDI Velocity in Automatic Piano Transcription},
  pdfauthor = {Zhanhong He, Roberto Togneri, Defeng (David) Huang}
}

\begin{document}
\capstartfalse
\maketitle
\capstarttrue

\begin{abstract}
MIDI velocity is crucial for capturing expressive dynamics in human performances. In practical scenarios, a music score with inaccurate velocities may be available alongside the performance audio (e.g., music education and free online archives), enabling the task of score-informed MIDI velocity estimation. In this work, we propose a modular, lightweight score-informed Transformer correction module that refines the velocity estimates of Automatic Music Transcription (AMT) systems. Leveraging high-quality expressive datasets for rigorous evaluation, we focus on piano transcription, integrating our module into multiple existing systems. Trained exclusively on the MAESTRO dataset, our method consistently reduces velocity errors and improves cross-dataset generalization to SMD and MAPS datasets. Under this training protocol, integrating our score-informed module with HPT (named Score-HPT) establishes a new state-of-the-art, outperforming existing score-informed methods and piano transcription models while adding only 1 M parameters. Furthermore, our framework remains robust to temporal misalignments caused by imperfect audio-score synchronization, a common real-world challenge.
\end{abstract}

\section{Introduction}\label{sec:introduction}
Automatic music transcription (AMT) is a longstanding topic in Music Information Retrieval (MIR), aiming to capture symbolic music information from audio. While traditional AMT systems focus on pitch and note boundaries to construct the basic MIDI score \cite{amt2013ov}, recent advancements have expanded to estimating MIDI velocity in piano performances \cite{amt2019ov}. This expansion has facilitated the creation of large-scale piano datasets \cite{hung2021emopia, kong2022giantMIDI, zhang2022atepp, edwards2023pijama, chou2024midibert}, benefiting downstream MIR tasks such as expressive rendering, music education and generation \cite{zhang2024dexter, he2025filling}. Despite recent progress, AMT outputs still require post-correction. Prior work has addressed temporal discrepancies through audio-to-score alignment \cite{morsi2022bottlenecks, zeit2024align, chang2025rumaa}. By contrast, our work focuses on refining the MIDI velocity predicted by AMT systems.

% jeong2019virtuosonet, rhyu2022sketching, tang2023reconstructing, borovik2023scoreperformer - expressiveness

Leveraging music score to aid velocity estimation from audio is known as score-informed MIDI velocity estimation. This paradigm is a practical scenario where a reference score and performance audio co-exist. In music education, for instance, students practice from a given score but require nuanced velocity feedback \cite{kim2022piano, kim2024method}. Besides, automated dataset curation often relies on online archives that provide aligned audio-score pairs but lack detailed dynamic annotations (e.g., MuseScore.com).

\begin{figure}[htbp]
\centering
\includegraphics[width=0.44\textwidth]{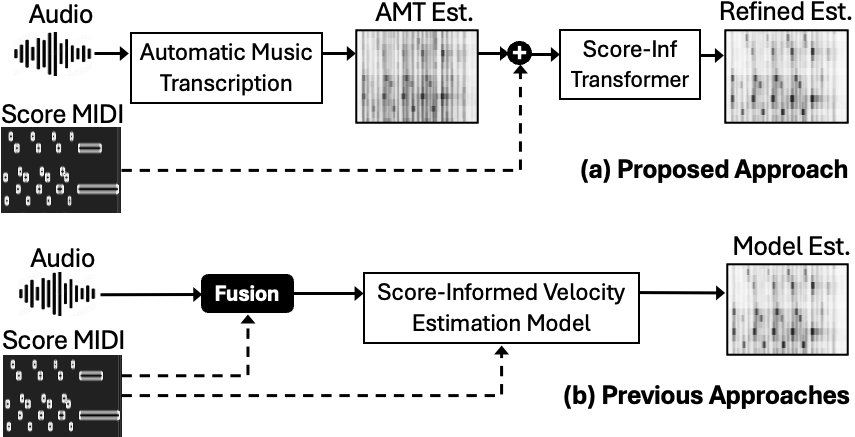}
\caption{Comparison between the proposed and previous score-informed MIDI velocity estimation approaches.}
\label{fig1}
\end{figure}

Previous score-informed approaches \cite{kim2023score, kim2024method} applied a FiLM conditioning layer to inject score information during audio feature processing. These feature fusion methods require the design of specific model architectures from scratch. In contrast, we propose a modular approach built upon existing piano AMT systems. Fig.~\ref{fig1} conceptually illustrates this distinction. Specifically, we introduce a lightweight Transformer encoder as a score-informed correction module. Inspired by how musicians reference a music score to make corrections, our design refines AMT-estimated MIDI velocity with minimal architectural changes, enabling easy integration into existing systems.

To evaluate its effectiveness, we integrated the proposed module into several well-known piano transcription models, including High-Resolution Piano Transcription (HPT) \cite{kong2021hpt}, HPPNet \cite{wei2022hppnet}, and a recent piano dynamic estimation (DynEst) system \cite{he2026dynest}. Furthermore, we conducted robustness evaluations under imperfect audio-score synchronization, demonstrating that our framework remains robust to temporal misalignments commonly encountered in real-world scenarios.\footnote{Code and pretrained models are available at: \\ \url{https://github.com/zhanh-he/score-informed-amt}}

\section{Related Work}

\begin{figure*}[ht]
\begin{center}
\includegraphics[width=0.98\textwidth]{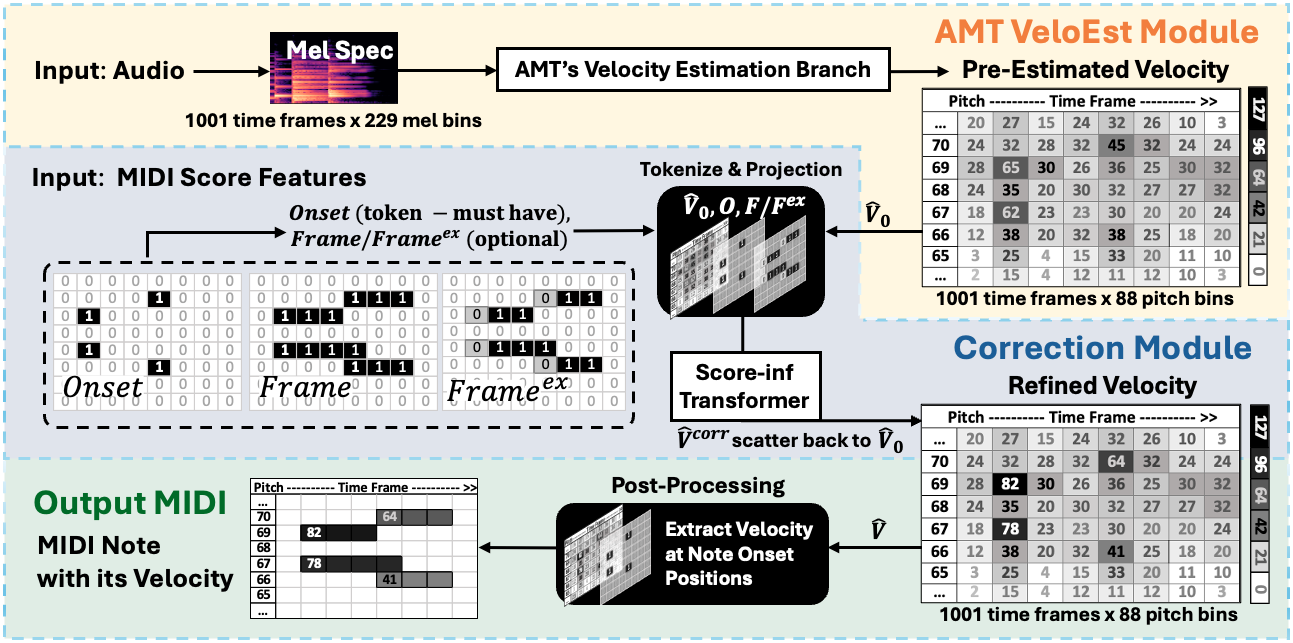}
\caption{Proposed score-informed MIDI velocity estimation framework. The system includes a velocity estimation module, which is identical to the velocity branch of the piano AMT system. The correction module then rectifies the preliminary velocity estimates using features extracted from the MIDI score.}
\label{fig2}
\end{center}
\end{figure*}

\subsection{Score-Informed MIDI Velocity Estimation}
Early AMT systems often ignored MIDI velocity \cite{amt2013ov}, fostering score-informed velocity estimation as a parallel field. Initial methods relied on manual sound pressure measurements and statistical analysis \cite{dan2006psy, goebl2003role, szeto2005finding}, leading to the first parametric estimation system in 2011 \cite{ewert2011est}. Subsequent research using restricted Boltzmann machines \cite{van2014pred} and non-negative matrix factorization \cite{jeong2017note, jeong2018timbre} proved the feasibility of machine learning for this task. However, their reliance on expert-defined parameters limited generalization and practical deployment.

Recently, deep learning approaches for score-informed velocity estimation have been introduced \cite{simonetta2022acoustics, kim2023diffvel, kim2023score}, with the FiLM U-Net model \cite{kim2024method} achieving state-of-the-art (SOTA) results. While these end-to-end designs eliminate the need for manual parameter tuning, contemporary audio-only AMT systems have also achieved competitive velocity estimation performance. This convergence motivates our approach: rather than developing a standalone score-informed model from scratch, we propose refining the initial velocity estimates from existing piano AMT systems by leveraging available score information.

\subsection{Automatic Piano Transcription}
Recent piano AMT systems have increasingly been estimating MIDI velocity, thanks to high-quality expressive piano datasets~\cite{maestro,smd,maps}. This progression includes the Onsets and Frames \cite{hawthorne2018onf}, the T5 Transformer \cite{hawthorne2021trans}, the HPT system \cite{kong2021hpt}, and the Hierarchical Frequency-Time Transformer (hFT-Transformer) \cite{toyama2023hft}. With a different focus, Semi-CRF \cite{yan2021semicrf} was proposed to refine pitch and note-boundary estimates, while its recent development using Transformer (Transkun v2) \cite{yan2024semicrf} achieved significant Notew/Off\&Velo scores across many piano datasets, representing the current SOTA. Meanwhile, other AMT studies propose competing solutions that prioritize broader generalization \cite{maman2022general, edwards2024general, strahl2024semi}, parameter efficiency \cite{wei2022hppnet}, or extend the focus to pedal \cite{wang2025multitask}.

Among these systems, HPT serves as a highly prevalent baseline. It has been used in creating well-known piano datasets (e.g., EMOPIA \cite{hung2021emopia}, GiantMIDI \cite{kong2022giantMIDI}, ATEPP \cite{zhang2022atepp}, PiJAMA \cite{edwards2023pijama}, Pianist8 \cite{chou2024midibert}) and has been adapted for guitar transcription \cite{riley2024high, riley2024gaps} as well as downstream MIR tasks. While HPT may not hold the current SOTA for piano transcription accuracy, its lightweight, modular CRNN architecture ensures computational efficiency and ease of modification. Similar advantageous qualities are found in HPPNet \cite{wei2022hppnet} and DynEst \cite{he2026dynest}, making them ideal candidates for evaluating our proposed approach.

% \cite{chou2024midibert}, music synthesis and source separation \cite{lin2021unified, cheuk2023jointist}.

\section{Methodology}

\subsection{MIDI Score Features}
\label{sec:score_feat}
Consistent with the standard paradigm of score-informed MIDI velocity estimation \cite{kim2024method}, we utilize MIDI scores that are time-aligned with the audio. We rasterize the score into pianoroll-like binary matrices
at the same time resolution as the AMT output. Let $T$ be the number of frames and $P$ the number of piano keys ($T{=}1001$ and $P{=}88$ in our setting). We define:
(i) onset matrix $\mathbf{O}\in\{0,1\}^{T\times P}$,
(ii) frame matrix $\mathbf{F}\in\{0,1\}^{T\times P}$,
and (iii) onset-excluded sustain matrix $\mathbf{F}^{\mathrm{ex}}\in\{0,1\}^{T\times P}$:
\begin{equation}
\mathbf{F}^{\mathrm{ex}} = \mathbf{F} - \mathbf{O}.
\end{equation}
We further extract the nonzero entries of these matrices:
\begin{equation}
\mathcal{I}=\{(t,p)\mid \mathbf{O}_{t,p}=1\},\quad
\mathcal{J}=\{(t,p)\mid \mathbf{F}_{t,p}=1\}.
\end{equation}
They are used for tokenization in our score-informed correction module (Sec.~\ref{sec:arch}) and masking the loss function.

\subsection{Model Architecture}
\label{sec:arch}
The proposed architecture, illustrated in Fig.~\ref{fig2}, adopts a modular design. Since AMT systems are multi-task architectures that estimate onset, offset, frame, and velocity simultaneously, we isolate their velocity estimation branch and extend it with our score-informed correction module.

\textbf{Velocity Estimation Module:}
This component is identical to the velocity branches of the selected piano AMT systems, including HPT \cite{kong2021hpt}, HPPNet \cite{wei2022hppnet}, and DynEst \cite{he2026dynest}. All setups follow the original implementations and their public repositories. We unify the input for all baselines using log-Mel spectrograms (10 second audio segmentation, waveform normalized, 22.05\,kHz sampling rate, 2048 FFT size, 100\,FPS, Hann window, 229 mel bins), producing an input tensor of $1001 \times 229$. The velocity branch outputs a preliminary velocity map $\hat{\mathbf{V}}_{0}\in[0,1]^{T\times P}$.

\begin{figure}[h]
\begin{center}
\includegraphics[width=0.48\textwidth]{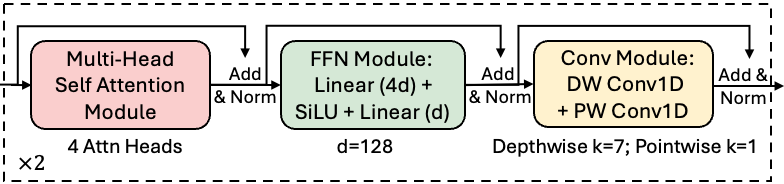}
\caption{Proposed \emph{Conformer-like} Transformer encoder.}
\label{fig3}
\end{center}
\end{figure}

\textbf{Score-informed Correction Module:}
The preliminary velocity $\hat{\mathbf{V}}_{0}$ is refined using the onset as tokens.
For computational efficiency, we tokenize only the score onset coordinates $\mathcal{I}$,
resulting in a variable-length sequence of $N=|\mathcal{I}|$ tokens per segment.
For each onset token $i=(t_i,p_i)\in\mathcal{I}$, we form an embedding
\begin{equation}
\mathbf{x}_i
= \mathbf{e}^{(t)}_{t_i} + \mathbf{e}^{(p)}_{p_i} + \mathbf{W}_f\,\mathbf{f}_i,
\end{equation}
where $\mathbf{e}^{(t)}_{t_i}$ and $\mathbf{e}^{(p)}_{p_i}$ are learnable pitch/time embeddings. The $\mathbf{f}_i$ contains the onset-wise velocity $\hat{V}_{0,t_i,p_i}$ and optional score-derived features ($F$ or $F^{ex}$), projected by each component's weight $\mathbf{W}_f$. Stacking all embeddings yields $\mathbf{X}\in\mathbb{R}^{N\times d}$, which is processed by the proposed lightweight Transformer encoder (a streamlined variant of the Conformer architecture~\cite{gulati2020conformer}), as detailed in Fig.~\ref{fig3}.
We use $d{=}128$, $2$ layers, and $4$ attention heads, applying a dropout probability of $0.1$ after each sub-module. Given the encoder outputs, a linear head predicts a scalar residual $\delta_i$ for each token, and we apply a bounded update:
\begin{equation}
\hat{V}^{\mathrm{corr}}_{t_i,p_i}
=\operatorname{clip}\!\Big(\hat{V}_{0,t_i,p_i}+\alpha\tanh(\delta_i),\,0,\,1\Big),
\end{equation}
where $\alpha$ controls the maximum correction magnitude (empirically set to $\alpha{=}0.2$).
The refined values replace the preliminary estimates in $\hat{\mathbf{V}}_{0}$ at onset positions, while non-onset entries remain unchanged.

\subsection{Loss Function}
\label{sec:loss}
All models are trained with the hybrid loss proposed in \cite{kim2024method}. It combines frame-wise BCE with onset-wise $\ell_1$:
\begin{equation}
\mathcal{L}
=\frac{1}{|\mathcal{J}|}\sum_{(t,p)\in\mathcal{J}}
\ell_{\mathrm{bce}}(V_{t,p},\hat{V}_{t,p})
+
\frac{1}{|\mathcal{I}|}\sum_{(t,p)\in\mathcal{I}}
\left|V_{t,p}-\hat{V}_{t,p}\right|.
\label{eq:loss}
\end{equation}
where $\mathbf{V}$ and $\hat{\mathbf{V}}$ denote ground-truth and predicted velocity of either (i) the AMT velocity branch or (ii) the same branch with the proposed score-informed module applied. In our preliminary experiments, this multi-scale supervision consistently outperformed single-target alternatives (e.g., frame-only or onset-only losses). Consequently, we adopt this formulation throughout our study, omitting detailed ablation results for brevity.

\subsection{Post-Processing}
\label{sec:post}
Velocity is defined at note onset. It reflects the instantaneous intensity of a keystroke rather than a time-varying loudness curve \cite{dan2006psy}. This onset-centered definition is consistent with the physical mechanism of the piano. Once a key has been fully pressed down, subsequent changes in finger pressure do not affect the tone. In other words, vibrato is technically impossible on an acoustic piano.

In our setting, a time-aligned MIDI score is available, but it does not provide reliable expressive dynamics labels such as note velocity. Instead, it only provides the ground-truth note positions, i.e., onsets and frames. We therefore bypass AMT-style note detection and focus only on velocity estimation. Consequently, we take the predicted velocity at each onset and assign this value to the entire note event, consistent with the MIDI specification.

\section{Experiment}

\subsection{Dataset}
In this study, we used the MAESTRO v3.0.0 dataset \cite{maestro} with its default train/validation/test split. This dataset comprises 1,276 Yamaha Disklavier piano performances, totaling over 200 hours of precisely aligned audio-MIDI data. A Yamaha Disklavier is an acoustic grand piano with an integrated electronic system that records MIDI data from human performance. Thus, MAESTRO provides audio captured by microphones, reflecting the acoustic environment, and its tightly synchronized MIDI data.

Following previous studies \cite{kim2024method}, we trained our model exclusively on the MAESTRO train set, and evaluated it on the Saarland Music Data (SMD) dataset \cite{smd}. SMD consists of human performances recorded on a Yamaha Disklavier piano, but differs from MAESTRO in acoustic environment and recording conditions. To further assess cross-dataset generalization, we evaluated our model on 60 piano recordings from the MAPS dataset \cite{maps}, exposing it to an additional set of acoustic scenarios.

\subsection{Training Setup}
For each selected AMT system, we trained two models in similar configurations. The baseline model is the standalone velocity branch of the original AMT system. For instance, our HPT baseline utilizes only the velocity branch, which contains about one-quarter of the full HPT system's parameters. Its score-informed counterpart (e.g., Score-HPT) integrates this same branch with our proposed correction module. This ensures a fair comparison by isolating the impact of score information on velocity estimation.

Each model was trained for 120k iterations on a single NVIDIA RTX 3090 (24\,GiB). We used Adam with learning rate $1\times10^{-4}$ (multiplied by $0.9$ every 10k iterations), batch size 12, and random seed 86. All models were trained on the MAESTRO training split. We selected the best checkpoint by validation set performance and then proceeded to evaluation.

% For each selected AMT model, we retrained its velocity branch as a baseline and compared it with a jointly trained counterpart augmented with the score-informed module applied. We optimize using the hybrid velocity loss proposed in \cite{kim2024method}, combining frame-wise BCE and onset-masked $\ell_1$:

\subsection{Evaluation Metrics}

Consistent with \cite{kim2024method}, we use the mean absolute error (MAE) and standard deviation of the error (STD) as evaluation metrics, calculated on the standard MIDI velocity scale of $[0,127]$. By applying onset-masking, we ensure a note-level evaluation that avoids the temporal bias of frame-wise metrics, where sustained notes would otherwise disproportionately influence the total error. These metrics are defined as:
\begin{equation}
\begin{aligned}
\mathrm{MAE} &= \frac{1}{|\mathcal{I}|} \sum_{i \in \mathcal{I}} \left| V_{i} - \hat{V}_{i} \right|, \\
\mathrm{STD} &= \sqrt{\frac{1}{|\mathcal{I}|} \sum_{i \in \mathcal{I}} \left( | V_{i} - \hat{V}_{i} | - \mathrm{MAE} \right)^2}.
\end{aligned}
\label{eq:eval}
\end{equation}

While previous work~\cite{kim2024method} (which predicts a sparse map) used Recall to penalize missing velocity at note positions, this metric is not discriminative in our setting. As shown in Fig.~\ref{fig4}, both the AMT baseline and our method output fully-populated velocity maps, so the velocity is non-zero at all potential note positions, leading to a Recall close to 100\%. Since the post-processing is applied to extract velocities at known note positions, we prioritize numerical precision (MAE / STD) rather than penalizing off-note predictions.

\section{Results and Discussion}

\subsection{Validation of Score-Informed Module}
To validate the proposed score-informed Transformer correction module, we first integrate it into the HPT \cite{kong2021hpt}. We choose HPT as an initial testbed because it is widely used and lightweight, which allows us to
isolate the contribution of the correction module before extending to additional baselines. We denote the resulting system
as Score-HPT.

\begin{table}[b]
    \centering
    % \resizebox{\columnwidth}{!}{
    \setlength{\tabcolsep}{4pt}
    \fontsize{10pt}{12pt}\selectfont
    \begin{tabular}{cll|ccc}
    \toprule
    \multicolumn{3}{c|}{\multirow{2}{*}{\textbf{Model \& Inputs}}}
          & \multicolumn{3}{c}{\textbf{MAESTRO test set}} \\
        \cmidrule(l){4-6}
          & & & MAE $\downarrow$ & STD $\downarrow$ & \# Params \\
        \midrule
        
    \multicolumn{2}{l}{\textbf{HPT}}          & \multicolumn{1}{l|}{} &  &  &  \\
    audio  & \multicolumn{2}{l|}{}            & 4.6 & 6.1 & 4.1 M \\
    \midrule
    \multicolumn{2}{l}{\textbf{Score-HPT}}  & \multicolumn{1}{l|}{}   &  &  &  \\
    audio  & + onset  & {}                    & 4.0  & 4.8  & 5.0 M \\
    audio  & + onset  & + frame               & 4.0  & \textbf{4.7}  & 5.0 M \\
    audio  & + onset  & + frame$_{\text{ex}}$ & \textbf{3.9}  & \textbf{4.7}  & 5.0 M \\
    \bottomrule
    \end{tabular}
    \caption{Comparison of HPT and Score-HPT on the MAESTRO test set. \(\downarrow\) indicate a lower value is better.} 
    % }
    \label{table1}
\end{table}

\begin{figure}[t]
\centering
\includegraphics[width=0.47\textwidth]{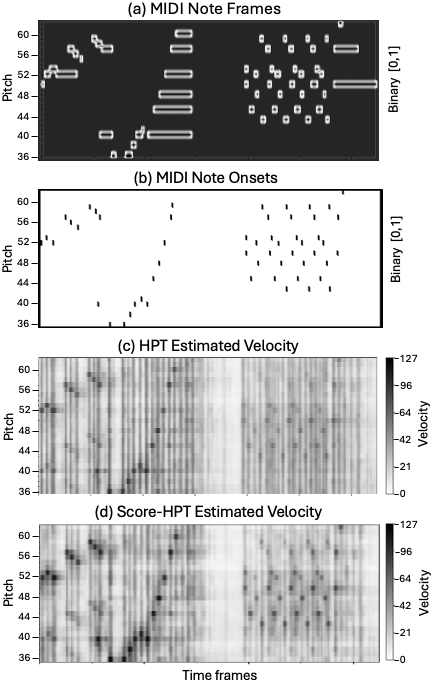}
\caption{Velocity estimates from the HPT baseline and the proposed Score-HPT. Referencing the given MIDI note features, the refinement effect is evident. Both HPT and Score-HPT provide a fully-populated velocity estimation.}
\label{fig4}
\end{figure}

Table~\ref{table1} summarizes the results on the MAESTRO test set. Compared with the original HPT, Score-HPT consistently achieves lower MAE and STD under all conditioning variants, with only a modest increase in model parameters (from 4.1\,M to 5.0\,M). Among these Score-HPT variants, \textit{onset + frame$_{\text{ex}}$} yields the best overall performance, while \textit{onset + frame} performs comparably. These results confirm that the proposed score-informed module effectively refines velocity estimates upon an existing AMT system. Additionally, Fig.~\ref{fig4} provides a qualitative visualization of this improvement. Compared to the HPT, Score-HPT generates velocity estimates that are more closely aligned with the MIDI note onset positions, clearly demonstrating the guiding effectiveness of the proposed score-informed Transformer module.

\begin{figure*}[ht]
\begin{center}
\includegraphics[width=0.98\textwidth]{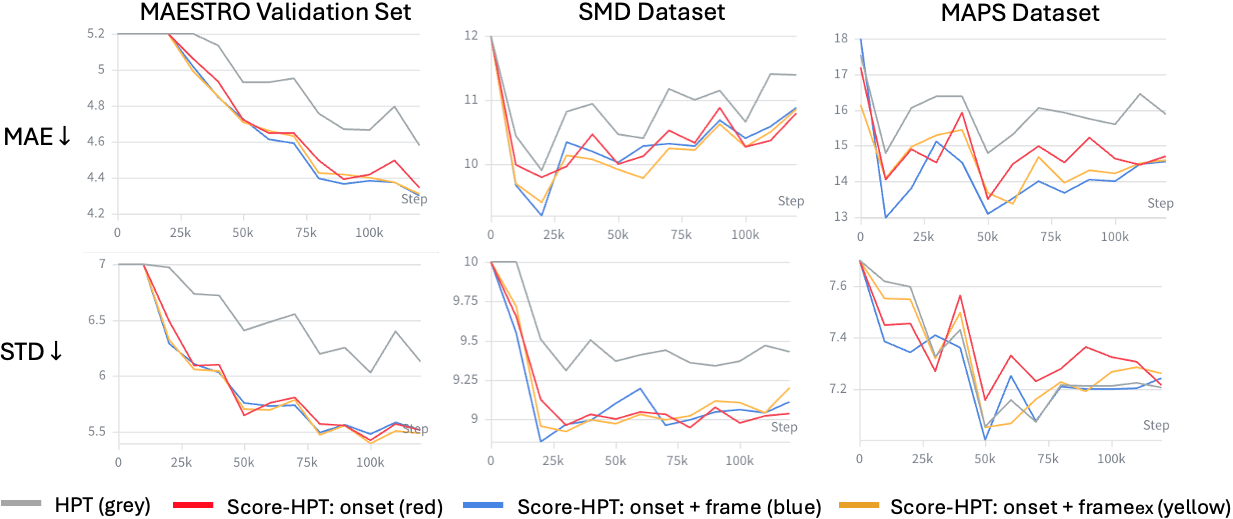}
\caption{Comparison between HPT and Score-HPT variants using different score features. Models were trained on the MAESTRO train set for 120k iterations and evaluated every 10k iterations across three validation sets.}
\label{fig5}
\end{center}
\end{figure*}

\subsection{Generalization across Datasets}
\label{sec:general}
Fig.~\ref{fig5} illustrates the learning curves for the HPT and Score-HPT variants. While training on the MAESTRO train set, models were evaluated on MAESTRO validation set, alongside 10\% held-out subsets of SMD and MAPS datasets (which lack official validation splits). All variants of Score-HPT consistently outperform the HPT baseline on both in-domain (MAESTRO) and out-of-distribution (SMD, MAPS) data, demonstrating superior generalization. The \textit{onset + frame} condition yields the best overall performance, though differences between conditions are modest.

During training in MAESTRO, the in-domain performance steadily improves but out-of-distribution performance peaks early and declines. This indicates that prolonged training overfits to MAESTRO-specific acoustics and harms generalization. Similar sound biases in AMT velocity estimation were also reported in \cite{martak2026soundmusicbias}. Nevertheless, Score-HPT improves both in-domain and out-of-distribution results. This highlights the benefits of score information, which provides an acoustics-invariant structural cue for reliable velocity estimation across diverse environments.

\subsection{Comparison with Existing Works}
\label{sec:existing}
Table~\ref{table2} evaluates the proposed score-informed framework against existing methods and representative AMT systems on the SMD dataset. All models were trained on MAESTRO only, enabling a fair comparison. Following the findings in Sec.~\ref{sec:general}, we employ the \textit{onset + frame} configuration for all proposed variants.

\begin{table}[h]
    \centering
    \setlength{\tabcolsep}{5pt}
    \fontsize{10pt}{12pt}\selectfont
    \begin{tabular}{l|rrr}
    \toprule
    \multicolumn{1}{c|}{\multirow{2}{*}{\textbf{\fontsize{10pt}{10pt}\selectfont Model}}}
      & \multicolumn{3}{c}{\textbf{SMD dataset}} \\
    \cmidrule(l){2-4}
      & MAE $\downarrow$
      & STD $\downarrow$
      & \# Params \\
    \midrule
    
    \textbf{Score-Inf. Methods} & & & \\
    DiffVel \cite{kim2023diffvel}         
        & 19.7 & 13.1 & -- \\
    FiLM Conv \cite{kim2023score}         
        & 15.1 & 12.3 & -- \\
    FiLM U-Net \cite{kim2024method}       
        & 9.9 & 7.8 & 8.8 M \\    
    \midrule
    
    \textbf{Proposed} & & & \\
    Score-HPT     & \textbf{8.0} & \textbf{6.9} & 5.0 M \\
    Score-HPPNet  & 9.2          & 7.3          & 1.5 M \\
    Score-DynEst  & \textbf{8.0} & \textbf{6.9} & 14.3 M \\
    \midrule

    \textbf{AMT Systems} & & & \\
    HPT \cite{kong2021hpt}                 & 10.0 & 8.8 & 4.1 M \\
    HPPNet (no CQT) \cite{wei2022hppnet}   & 10.6 & 9.0 & 0.5 M \\
    DynEst (no BSSL) \cite{he2026dynest}   & 10.4 & 8.8 & 13.3 M \\
    hFT-Transformer \cite{toyama2023hft}   & 9.9  & 7.3 & 5.5 M \\
    Transkun v2 \cite{yan2024semicrf}      & 13.3 & 11.9 & 13.6 M \\
    \bottomrule
    
    \end{tabular}
    \caption{Comparison of different methods on the SMD dataset.
    $\downarrow$ indicate a lower value is better.}
    \label{table2}
\end{table}

To further demonstrate the generality of our approach, we integrated the correction module into diverse architectures beyond HPT \cite{kong2021hpt}, including HPPNet \cite{wei2022hppnet} and DynEst \cite{he2026dynest}. Although retraining these models on log-Mel spectrograms rather than their original CQT or Bark-scale specific loudness (BSSL) features inherently degrades their performance, our score-informed module consistently yields significant improvements. This confirms that our approach serves as a general-purpose enhancement for different piano AMT backbones, while adding only a marginal overhead of $\sim$1M parameters. Through this integration, Score-HPT and Score-DynEst achieve new SOTA performance. These models outperform both dedicated score-informed frameworks, such as FiLM U-Net \cite{kim2024method}, and top-tier AMT systems, including hFT-Transformer \cite{toyama2023hft} and Transkun v2 \cite{yan2024semicrf}. Interestingly, although Transkun v2 \cite{yan2024semicrf} currently defines the SOTA in AMT, it lags behind several other models in velocity MAE and STD. A plausible explanation is that its core SemiCRF is primarily designed for pitch and note onset/offset modeling, which enables strong transcription performance (Notew/Off\&Velo), whereas velocity estimation is not its main strength. Incorporating our score-informed correction module into Transkun v2 therefore remains a promising future direction.

\subsection{Robustness to Audio-Score Misalignment}
\label{sec:unaligned_eval}

\begin{figure}[t]
\centering
\includegraphics[width=0.49\textwidth]{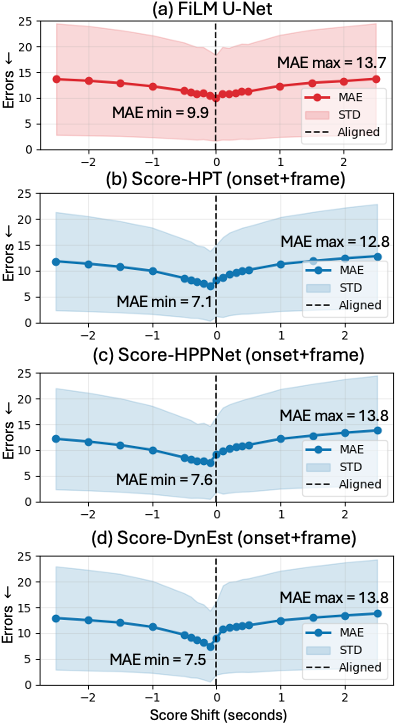}
\caption{Robustness evaluation on data derived from the SMD dataset, using audio-score misalignments of $\pm$0.1, 0.2, 0.3, 0.4, 0.5, 1, 1.5, 2, and 2.5 s.}
\label{fig6}
\end{figure}

To evaluate the models against audio-score temporal misalignments, we conducted robustness tests using timing shifts ranging from $\pm$0.1 to $\pm$2.5 seconds, following \cite{kim2024method}. We applied these artificial shifts on SMD dataset, forming the evaluation data. For example, a $+0.1$ shift means that the original, perfectly aligned score is intentionally delayed by 0.1 seconds. All models evaluated here are exactly the same as those in Sec.~\ref{sec:existing}, meaning they were trained exclusively on the MAESTRO dataset (not shifted, perfect synchronization). The final robustness results are summarized in Fig.~\ref{fig6}.

Specifically, we introduce a necessary refinement to the evaluation pipeline of \cite{kim2024method}. Their approach used shifted scores to feed the model but unshifted scores for post-processing to be consistent with the earlier benchmark \cite{kim2023score}. While this intention is understandable, it introduces a timeline mismatch: if a model is guided by a shifted score, its predictions are inherently tied to that shifted temporal grid. Therefore, we argue that the model's output should be post-processed using the shifted MIDI score. Under this more internally consistent protocol, FiLM-UNet performs significantly better than originally reported in \cite{kim2024method}, revealing its true robustness as shown in Fig.~\ref{fig6}.

Applying this refined evaluation protocol, Score-HPT achieves the best performance on the SMD dataset under misaligned settings. Notably, all methods in Fig.~\ref{fig6} degrade only marginally compared to the audio-score perfectly aligned scenario (score shift $=$ 0 second). Given that MIDI velocity ranges from 0 to 127, an increase of around 6 units in MAE or STD under misalignment corresponds to a very small absolute deviation relative to the full velocity range. Furthermore, we observe that minor shifts unexpectedly improve the performance of our proposed score-informed variants, likely by offsetting the inherent fuzziness of the AMT branch's velocity prediction. Together, these findings highlight an important conclusion: score-informed MIDI velocity estimation methods are highly robust to audio-score misalignments.

\section{Conclusion and Future Work}
In this work, we presented a novel, modular score-informed Transformer correction module for refining MIDI velocity in automatic piano transcription when a reference score is available. By operating on score onset tokens and predicting bounded residual updates, the proposed module can be attached to an existing AMT velocity branch with minimal architectural changes. Experiments with MAESTRO-only training demonstrate consistent improvements across multiple piano transcription systems (HPT, HPPNet, and DynEst), stronger cross-dataset generalization to SMD and MAPS, and state-of-the-art performance on SMD under our training protocol, surpassing both prior score-informed models and competitive piano transcription systems. We further demonstrate that the proposed framework is robust to audio-score temporal misalignments.

A key limitation of the current study is its restriction to piano datasets, constrained by the fact that high-fidelity MIDI velocity ground truth is predominantly captured via acoustic pianos equipped with velocity sensors. For many other instruments, velocity annotations are scarce or unavailable, making supervised training and evaluation difficult. An important direction for future work is to transfer the proposed framework from piano to other instruments for MIDI velocity estimation, potentially leveraging self-supervised learning, transfer learning, or continual learning to reduce dependence on explicit velocity labels.

\begin{acknowledgments}
This research was carried out while the author was in receipt of a University International Fees Scholarship and a University Postgraduate Award at The University of Western Australia. The authors acknowledge the use of Kaya, the High-Performance Computing facility at The University of Western Australia, and the technical assistance provided by the UWA IT -- HPC team.
\end{acknowledgments}
	
%%%%%%%%%%%%%%%%%%%%%%%%%%%%%%%%%%%%%%%%%%%%%%%%%%%%%%%%%%%%%%%%%%%%%%%%%%%%%
%bibliography here
\bibliography{smc2026}
	
\end{document}